# Linear simulation of ion temperature gradient driven instabilities in W7-X and LHD stellarators using GTC

*Hongyu Wang,[1] Zhihong Lin,[1,2] Ihor Holod,[2] Jian Bao,[1,2] Lei Shi,[2] and Sam Taimourzadeh[2]*

[1]*Fusion Simulation Center, Peking University, Beijing 100871, China*

[2]*Department of Physics and Astronomy, University of California, Irvine, California 92697, USA*

Abstract

The global gyrokinetic toroidal code (GTC) has been recently upgraded to do simulations in non-axisymmetric equilibrium configurations, such as stellarators. Linear simulations of ion temperature gradient driven instabilities (ITGs) have been done for Wendelstein7-X (W7-X) and Large Helical Device (LHD) stellarators using GTC. The capability of capturing the main characteristics of ITGs in stellarators are demonstrated by several numerical measures, including the convergence of real frequencies, growth rates, and parallel spectrum in parallel resolution and toroidal mode number filtering, the poloidal and parallel spectrums, and the electrostatic potential mode structure on a flux surface.

## I. INTRODUCTION

After stellarators were designed in 1950s, several concepts in magnetic confinement devices are first presented in stellarators [reference1] and recently increasing attentions have been put on magnetic confinement devices with symmetry-breaking effects [reference2]. The non-axisymmetric feature of the stellarators enables many unique physical phenomena and possible methods to achieve nuclear fusion. At the same time, considering the similarities of tokamaks and stellarators, like neo-classical transport and Alfven eigenmodes, research in stellarator physics also helps understand tokamaks better [reference 3]. Several experimental results in stellarators have been received. W7-X, the biggest stellarator in the world currently, first produced helium plasma in December 2015 [reference4]. LHD firstly produced plasma in 1998 [reference5 6] and nowadays plasma confinement properties of LHD are comparable to world's fusion devices. Characteristics of micro turbulence have been studied in H-mode plasma in LHD [reference7], and inward radial propagation of spontaneous toroidal flow have been observed in LHD [reference8].

As stellarator is one kind of hopeful devices to make nuclear fusion come true, it is necessary to study characteristics of plasma physics in stellarators in both experiments and computer simulations. Several efforts have been put into simulations on stellarators. EUTERPE, a global particle-in-cell code, has been used to study the effects of collisions on ITG in LHD [reference9] and radial electric fields effects on linear ITG in W7-X and LHD [reference10]. GENE has done nonlinear simulation of ITG turbulence in W7-X [reference11] and compared simulation results of microinstabilities and turbulence with EUTERPE [reference12]. Linear simulations of ITG instabilities have also been studied for both standard and inward-shifted LHD configurations using GKV code [reference13], and an upgraded code, GKV-X, has done linear simulations of ITG using precise magnetic configurations of LHD, and comparisons between linear simulation results of ITG using GKV-X and LHD experiments observations are also given [reference14].

Recently, the global gyrokinetic toroidal code (GTC) has been upgraded to use 3D equilibrium geometry. This enables the study of the plasma phenomena in stellarators caused by the non-axisymmetric equilibrium configurations. GTC is a global gyrokinetic particle-in-cell code [reference15], and uses the Boozer coordinates (ψ, θ, ζ) which is a specific magnetic flux coordinates system. The 3D equilibrium geometry is calculated by the Variational Moments Equilibrium Code (VMEC) [reference16], and loaded into GTC through a recently developed interface [reference 17 18].

In this paper, the ITGs in W7-X and LHD are studied using GTC and the simulations are done both in the full torus mode and the partial torus mode. The partial torus runs take advantage of the fact that the equilibrium geometries are periodic in toroidal direction, and have a period of 1/5 and 1/10 of the full torus in W7-X and LHD, respectively. Effects of the parallel resolution and the coupling between multiple toroidal modes on the growth rates and real frequencies are investigated. The electrostatic potential on a flux surface is also presented in both 2D and 3D to show both the similarities and the differences of ITGs in W7-X and LHD.

The remainder of this paper is structured as follows. The data interface between GTC and VMEC for the non-axisymmetric geometry is presented in Section II. In Section III, simulation results of ITGs in W7-X are given. Convergence in both the parallel resolution and the toroidal mode filtering is achieved, and unique features are observed and attributed to the field periodicity of W7-X in parallel direction. The electrostatic potential on a flux surface is plotted in a flat (θ, ζ) plane, as well as in a 3D view in real space to visualize the mode structure directly. Results of partial torus simulation in W7-X are also shown in this section. In section IV, the simulation results of ITGs in LHD are presented in the same structure. The convergence and mode structure studies show both similarities and differences between LHD and W7-X ITGs. Finally, conclusion and discussions are provided in Section V.

## II. CONSTRUCTION OF NON-AXISYMMETRIC MAGNETIC FIELDS FROM VMEC DATA

Thanks to the collaboration between the GTC team at UCI and the VMEC team at ORNL, GTC has recently been updated to treat 3D equilibria by interfacing with the MHD equilibrium code VMEC [reference 17 18]. The equilibrium geometry and magnetic field data from VMEC is provided in the form of Fourier series coefficients $B_{cn}$, $B_{sn}$ in the toroidal direction:

$$B(\psi,\theta,\zeta) = \sum_{n=1}^{N} \left[ B_{cn}(\psi,\theta,n)\cos(ntor(n)\zeta_n) + B_{sn}(\psi,\theta,n)\sin(ntor(n)\zeta_n) \right] \quad (1)$$

where (ψ, θ, ζ) are normalized poloidal flux, poloidal angle, and toroidal angle, respectively, forming right-handed Boozer coordinate system, N the total number of toroidal harmonics chosen, and `ntor(n)` the corresponding toroidal harmonic mode number. The Fourier coefficients $B_{cn}$ and $B_{sn}$ are specified on a rectangular mesh in the (ψ, θ) space, with the number of grid points (`lsp`, `lst`). The magnetic field's ζ-dependence is essential in the stellarators, and is taken into account in the particle trajectories, the gyrokinetic equations, and the electron continuity equation in GTC.

Similar to the magnetic field strength B, the cylindrical coordinates (X, Z) are also given in the form of Eq. (1). In addition, the flux functions representing the poloidal current, g(ψ), the toroidal current, I(ψ), the effective minor radius, r(ψ), and the magnetic safety factor, q(ψ), are all provided by VMEC.

We use quadratic spline interpolation along (ψ, θ, ζ) to efficiently evaluate the 3D equilibrium quantities in GTC. The spline coefficients are constructed on the identical mesh in ($\psi$, $\theta$) as the read-in VMEC data, and the same as the simulation grid along $\zeta$. The periodicity of the values and the first derivatives in $\theta$ and $\zeta$ directions is enforced by using periodic boundary condition when constructing the spline functions. An odd number of spline sections in these directions is required due to some intrinsic constraints in the periodic quadratic spline algorithm [reference19].

## III. SIMULATION OF ION TEMPERATURE GRADIENT DRIVEN INSTABILITIES IN W7-X

In a stellarator, neither the device configuration nor the produced plasma is axisymmetric. The shape of a flux surface changes along the toroidal direction. The poloidal contours of the flux surfaces look the same only after rotating the stellarator in toroidal direction by a certain angle. The field period, $N_{fp}$, is used to describe this discrete symmetry in stellarators, and the toroidal period of the field is simply given by $\frac{2\pi}{N_{fp}}$. For W7-X, $N_{fp}$ is 5, i.e. the magnetic field is symmetric under the rotation of 72° in toroidal direction. In Figure 1, the 2D poloidal contours of the diagnosis flux surface in W7-X are given at $\zeta = \frac{0}{N_{fp}}, \frac{0.5\pi}{N_{fp}}, \frac{\pi}{N_{fp}}, \frac{1.5\pi}{N_{fp}}$.

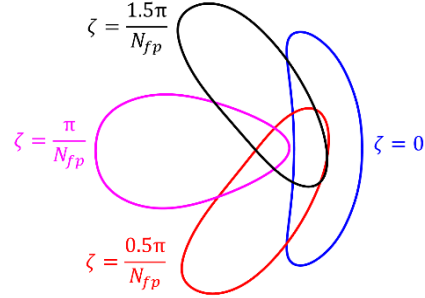

Figure 1. 2D poloidal contours of W7-X on diagnosis flux surface at $\zeta$ =0 (blue), $\zeta$ =0.5 $\pi$ / $N_{fp}$ (red), $\zeta$ = $\pi$ / $N_{fp}$ (magenta) and $\zeta$ =1.5 $\pi$ / $N_{fp}$ (black). The field period $N_{fp}$ =5 in W7-X.

The rotational transform, $\iota$, is used to describe the ratio between magnetic field line traveling angles in toroidal and poloidal directions in stellarators. Using tokamak terminology, $\iota$=1/q, where q is the safety factor commonly used in tokamaks. In our simulation, $\iota$ is 0.909 at the diagnosis point. The ion temperature profile, in which ion temperature gradient $\eta$ is 1, and the rotational transform profile used in our simulations are given in Figure 2(a). The simulation range of $\psi$ is also marked out in Figure 2, and the diagnosis flux surface is at $\psi$=0.5. Basic parameters used in the simulations of W7-X are as follows: magnetic field on axis is 23600 Gauss, ion temperature at diagnosis point is 1000 eV, major radius is 561.79 cm, $k_\perp\rho_i$=0.86, time step is 0.01 R0/Cs, and Cs/R0 is 2.43E-04 rad/s. Growth rates and real frequencies of the simulations in this section are normalized to Cs/R0. As introduced in Section II, Fourier series coefficients in $\zeta$ expansion are used to build the 3D magnetic field in GTC. Since W7-X has a field period of 5, the toroidal harmonic series `ntor(n)` is an arithmetic progression with common difference of 5, so the first 4 harmonics are 0, 5, 10, 15.

The first 4 Fourier series coefficients of the equilibrium magnetic field, evaluated at θ=0, $B_{cn}(\psi,1,1)$, $B_{cn}(\psi,1,2)$, $B_{cn}(\psi,1,3)$, $B_{cn}(\psi,1,4)$, are given in Figure 2(b) as functions of $\psi$.

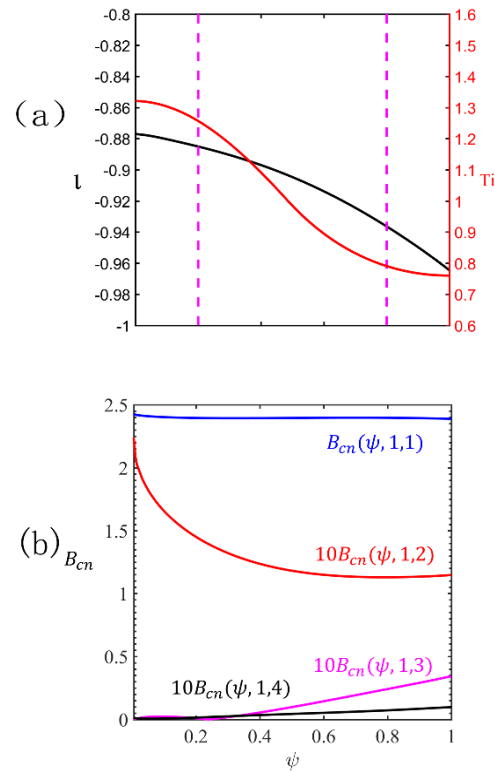

Figure 2. (a) Ion temperature (red) profile and rotational transform (black) profile. The simulation range is marked out by magenta lines: $\psi_{inner}$=0.2 and $\psi_{outer}$=0.8. (b) Fourier series coefficients of magnetic field of W7-X used in GTC, B$_{cn}$($\psi$,1,1)=0 (blue), B$_{cn}$($\psi$,1,2)=5 (red), B$_{cn}$($\psi$,1,3)=10 (magenta), B$_{cn}$($\psi$,1,4)=15 (black). Values of B$_{cn}$($\psi$,1,2), B$_{cn}$($\psi$,1,3), B$_{cn}$($\psi$,1,4) are multiplied by 10.

Considering the non-axisymmetric nature of the W7-X equilibrium, it is important to study the number of parallel grids needed in the simulations. The growth rates and real frequencies of `nmode=185`, `mmode=205` with parallel grids N$_p$ =17, 33, 65, and 97 are shown in Figure 3. The parallel spectrums for different parallel grids are given in Figure 4.

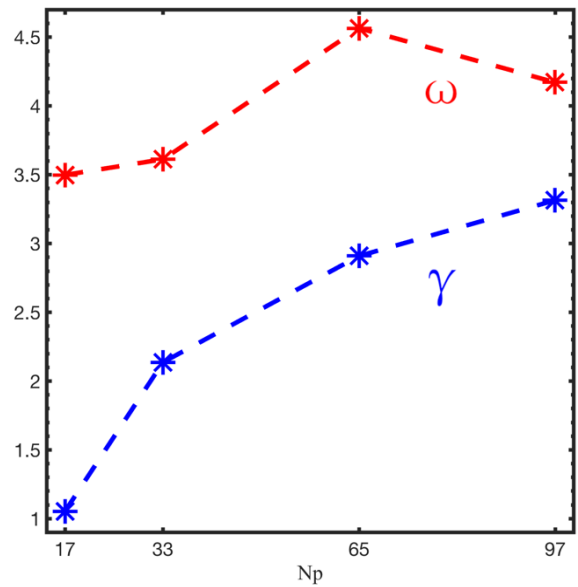

Figure 3. Growth rates (blue) and real frequencies (red) of parallel grids convergence.

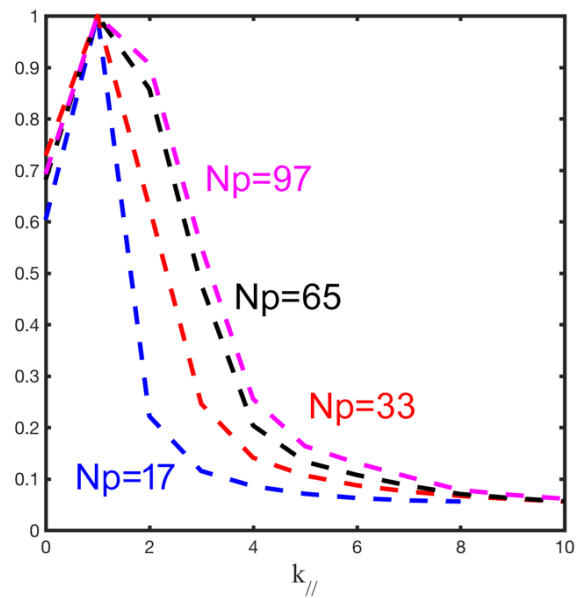

Figure 4. Parallel spectrums of different parallel grids.

According to the convergence shown in the growth rates, real frequencies and parallel spectrums, 65 parallel grids are needed to carry out full torus simulations in W7-X using GTC. Note that we have been able to resolve a toroidal mode structure as fine as nmode=185 with only 65 simulation planes. This is because the coordinate of GTC is field-aligned, and we only need to resolve the parallel structure of the perturbations. Thus the parallel grid number required for convergence is much fewer than that required to resolve the toroidal mode structure. This feature greatly saves the computing resources.

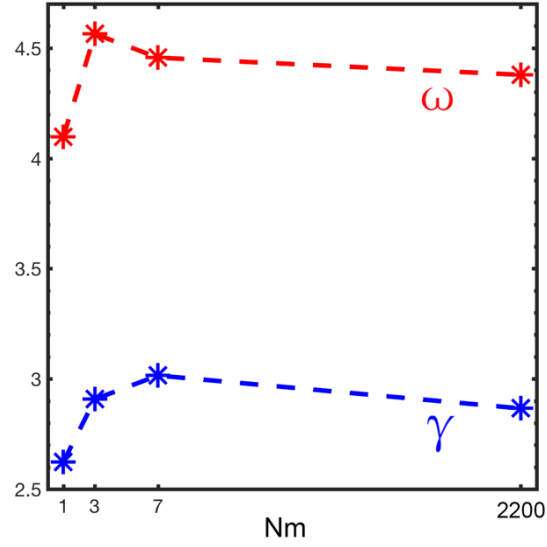

Figure 5. Growth rates (blue) and real frequencies (red) of nmode number convergence.

In GTC, after solving fields, a filtering step is put before gathering particles, so the number of the kept toroidal modes, `Nm`, can be set. To study different modes coupling effects in W7-X, we here show the convergence of `Nm` number. In 3 simulations with filtering, we have kept one, three, and seven toroidal modes, and compared the corresponding real frequencies and growth rates with the non-filtering case. In the one-toroidal-mode case, all modes except `nmode=185` are filtered out. Since the field period of W7-X is 5, the toroidal mode number must be integral multiplications of 5. So, `nmode=180, 185, 190` are kept when `Nm=3`, and `nmode=170, 175, 180, 185, 190, 195, 200` are kept when `Nm=7`. In the non-filtering case, the filtering step is thoroughly omitted. In the figures, this case is labeled with `Nm=2200`, which is the effective toroidal mode numbers given by the highest parallel resolution. Growth rates and real frequencies of `nmode=185` are shown in Figure 5. The growth rates and real frequencies of `Nm=3` and `Nm=2200` do not differ much, which indicates that `Nm=3` may be enough to be used in the simulations.

The growth rates and real frequencies of `nmode=185` at different ion temperature gradients η are given in Figure 6. Here, η is defined as $\eta \equiv -dlnTi/d\psi_{tor}$. In these simulations, `Nm=3` is used as discussed above. These growth rates and real frequencies show good linearity with ion temperature gradient and are compared with simulation results of EUTERPE.

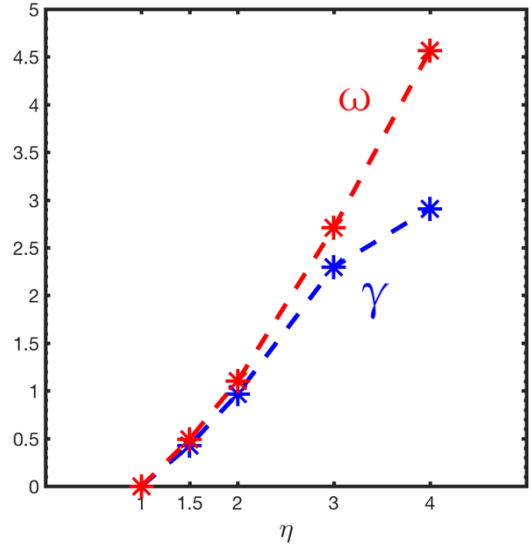

Figure 6. Growth rates (blue) and real frequencies (red) of nmode=185 and mmode=205 when keeping 3 nmodes in simulation.

Fast Fourier Transform Algorithm is used to calculate the poloidal spectrum on flux surfaces. An average poloidal spectrum of the diagnosis flux surface and 10 adjacent flux surfaces in the simulation, is given in Figure 7. The dominant mmode range of the poloidal spectrum appears from mmode=200 to mmode=600.

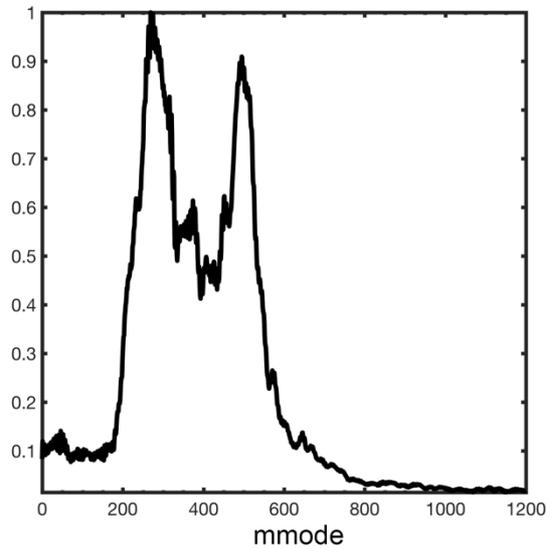

Figure 7. Averaging poloidal spectrum.

The electrostatic potential phi on the diagnosis flux surface is shown in Figure 8. The horizontal axis is the toroidal angle, $\zeta$, changing from 0 to $2\pi$, and the vertical axis is the poloidal angle in Boozer coordinates, $\theta$, changing from $-\pi$ to $\pi$. We can see 5 clusters in the ($\zeta$, $\theta$) space. The 3D mode structure of the electrostatic potential on the diagnosis flux surface is shown in Figure 9. Both

Figure 8 and Figure 9 show the discrete toroidal symmetry of W7-X and reflect the 5 field periods' effects. The mode structure is almost unchanged after rotating one field period, but the intensity of the potential is slightly different.

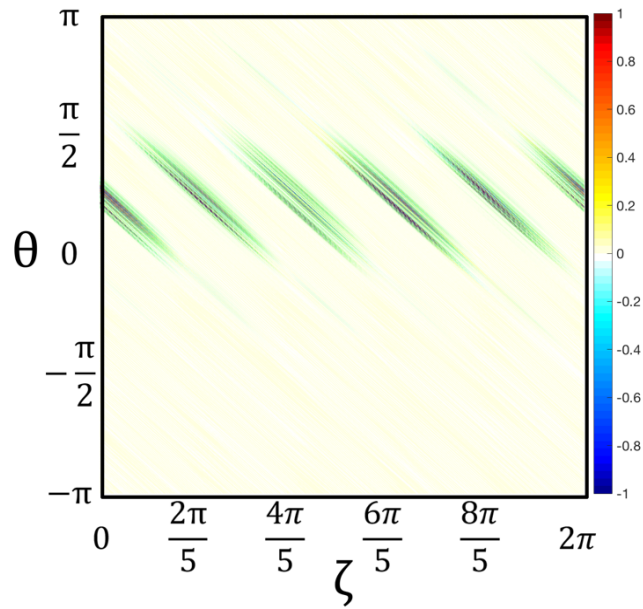

Figure 8. Phi on flux surface when keeping all nmodes. X axis is ζ and y axis is θ.

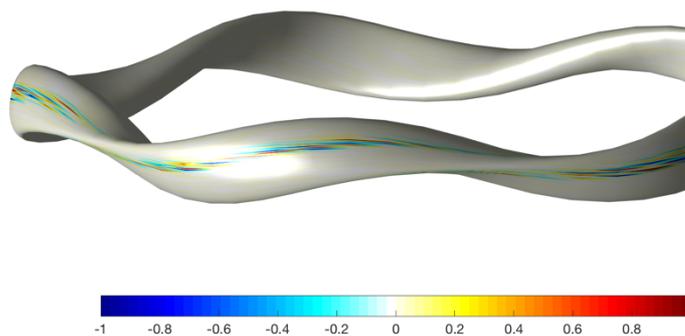

Figure 9. 3D electrostatic potential mode structure of W7-X.

ITG driven instability in W7-X is also studied in partial torus using GTC. Simulations of stellarators can be done in one field period as the periodic structure in toroidal direction of stellarators. One fifth of W7-X in toroidal direction, of which ζ is from 0 to 2π/5, is used in partial torus simulations. Parallel grids convergence is studied again in partial torus simulations of W7-X in Figure 10, and it shows that 25 parallel grids are needed in partial torus simulations for one field period, bigger than that in full torus simulations. Averaging poloidal spectrum of W7-X in Figure 11 is easily to be influenced by different parameters in simulations, and dominant mmode is not a certain value, and it is from mmode=200 to mmode=400. Compared with poloidal spectrum of full torus simulation, the unstable mmode range of partial torus is narrower. The electrostatic potential phi on the diagnosis flux surface in partial torus simulation of W7-X is shown in Figure 12 (left), and the horizontal axis is the toroidal angle, ζ, changing from 0 to 2π/5, and the vertical axis is the poloidal angle in Boozer coordinates, θ, changing from –π to π. According to mode structure of W7-X in 3D space in Figure 12(right), the mode structure also allocates at bad curvature range of W7-X and it is the same with that of full simulation, but the intensity of the electrostatic potential is almost the same at different clusters, which is different from that of full torus simulation. In partial torus simulations of W7-X, only particular nnodes, which are integers multiple of 5, are kept in simulations, while all nnmode are kept in full torus simulations, and that will cause the difference of mode structure on flux surface. Computing source can be saved by greatly decreasing parallel grids in partial torus simulations, and considering the difference of mode structure in partial torus simulation and full torus simulation, nmodes coupling is also proved to be important.

More grids in theta direction are used to cover higher mmode and study the relationship between growth rates, frequencies and eigenmodes. Growth rates and frequencies of different nmodes are given in Figure 13. $k_\theta \rho_i$ is proportional to nmode, and $k_\theta \rho_i = 0.86$ when nmode=185. Growth rates are almost the same from nmode=185 to nmode=360, while frequency is approximately proportional to nmode. Poloidal spectrum of this case is shown in Figure 14. As $k_\theta \rho_i$ is large in our case, influence of finite Larmor radius effects should be taken into consideration, which is called short wavelength ITG.[reference 20, 21, 22] Space Fourier series expansion is not applicable in stellarators analyzing, for symmetry breaking in both θ and ζ direction, and we can get $\phi(\psi, \theta, \zeta, t) = \sum \phi(\psi, \theta, \zeta) e^{-i\omega t}$, but space transform in ζ direction does not make sense on stellarators. Growth rates of nmodes are almost the same, because of the nmode coupling. In our analysis, we use filter to get growth rates and frequencies, so $\phi(\psi, \theta, \zeta, t) = e^{\gamma t} \sum_n \phi(\psi, \theta) e^{i(n\zeta - \omega t)}$, and frequency is approximately proportional to nmode.

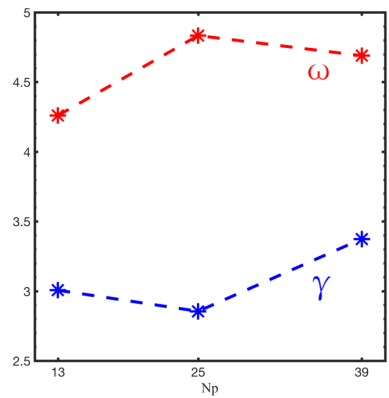

Figure 10. Growth rates (blue) and real frequencies (red) of parallel grids convergence in partial simulations of W7-X.

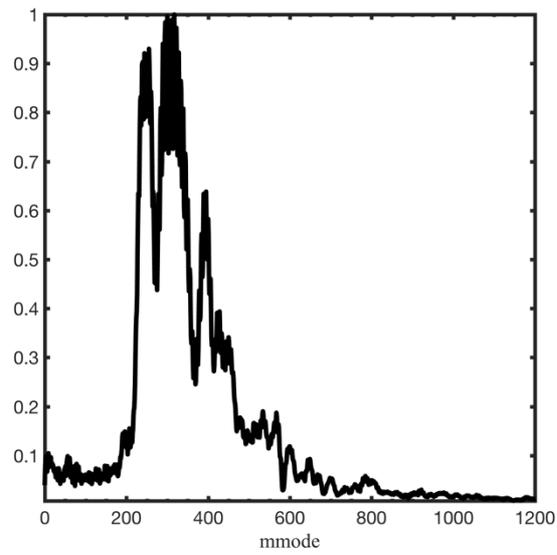

Figure 11. Averaging poloidal spectrum of partial torus simulation on W7-X.

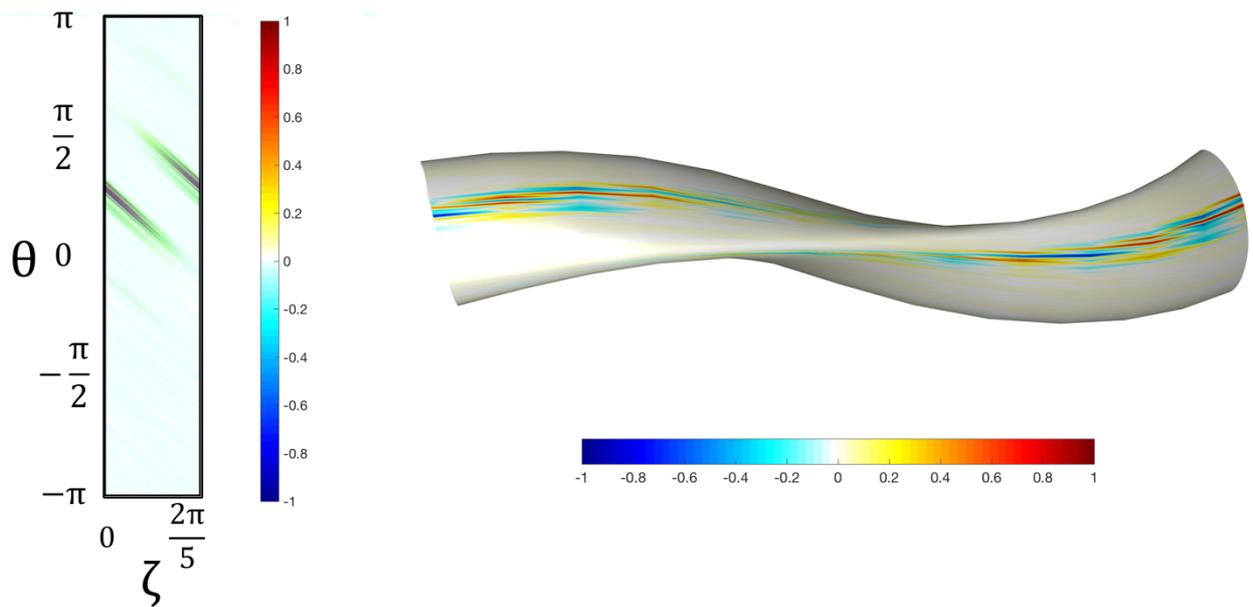

Figure 12. Electrostatic potential on flux surface (left) and 3D space (right) of one fifth of torus in toroidal direction of W7-X.

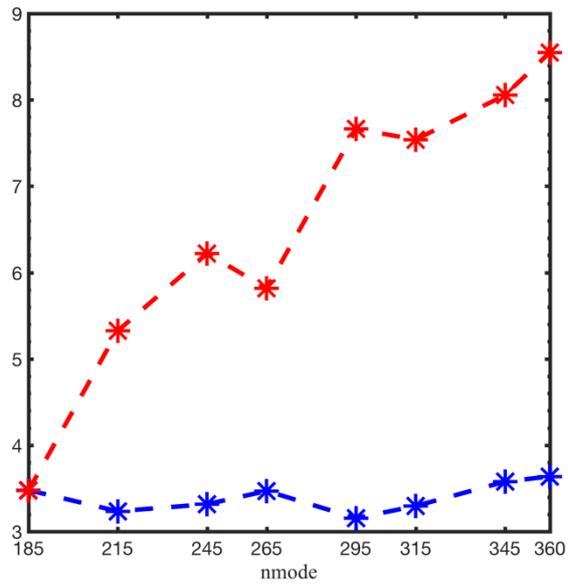

Figure 13. growth rates (blue) and frequencies (red) of different eigenmodes. X-axis is nmode.

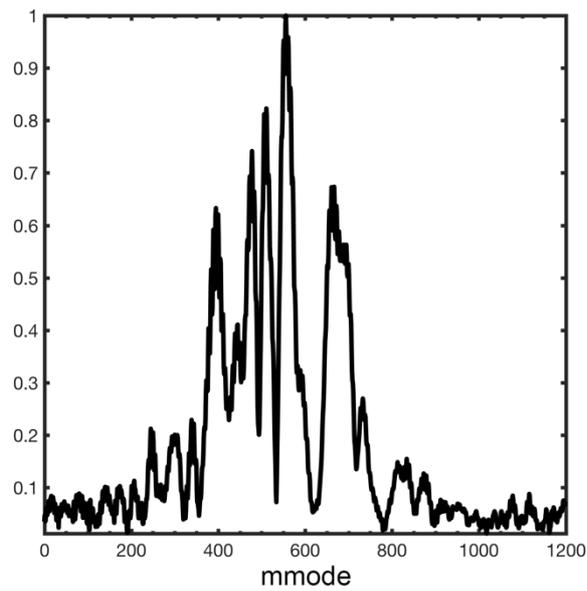

Figure 14. Poloidal spectrum of partial torus simulation on W7-X. More grids in theta direction are used in this case.

## IV. SIMULATION OF ITG IN LHD

In this section, we show the ITG simulation results in the LHD configuration, which has a field period of 10, $N_{fp}$=10. The 2D poloidal contours of the diagnosis flux surface at $\zeta = \dfrac{0}{N_{fp}}, \dfrac{0.5\pi}{N_{fp}}, \dfrac{\pi}{N_{fp}}, \dfrac{1.5\pi}{N_{fp}}$ are shown in Figure 15.

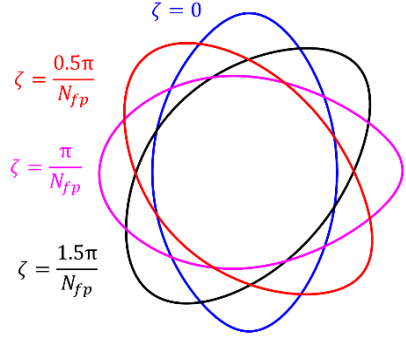

Figure 15. 2D poloidal contours of LHD on diagnosis flux surface at $\zeta$ =0 (blue), $\zeta$ =0.5 π / $N_{fp}$ (red), $\zeta$ = π / $N_{fp}$ (magenta) and $\zeta$ =1.5 π / $N_{fp}$ (black). $N_{fp}$ =10 in LHD.

The ion temperature and rotational transform profiles used in the simulations are presented in Figure 16(a), the simulation range of $\psi$ is indicated by the vertical dashed lines, and the diagnosis flux is at $\psi$=0.375. The basic parameters are as follows: the on-axis magnetic field is 14457.83 Gauss, the on-axis ion temperature is 1000 eV, the major radius is 372.8 cm, $k_\perp \rho_i$=0.40, time step is 0.004 R0/Cs, and Cs/R0 is 5.98E-04 rad/s. Growth rates and real frequencies are normalized to Cs/R0 in this section. Similar to the W7-X case, the 3D equilibrium in LHD is also provided by VMEC ithrough the Fourier series coefficients in $\zeta$. The first 4 Fourier series coefficients profiles $B_{cn}(\psi,1,1)$, $B_{cn}(\psi,1,2)$, $B_{cn}(\psi,1,3)$, $B_{cn}(\psi,1,4)$ used to construct the 3D magnetic field of LHD are given in Figure 16(b). Again, since the field period is 10 in LHD, the first 4 toroidal harmonics are `ntor(n)=0, 10, 20, 30`.

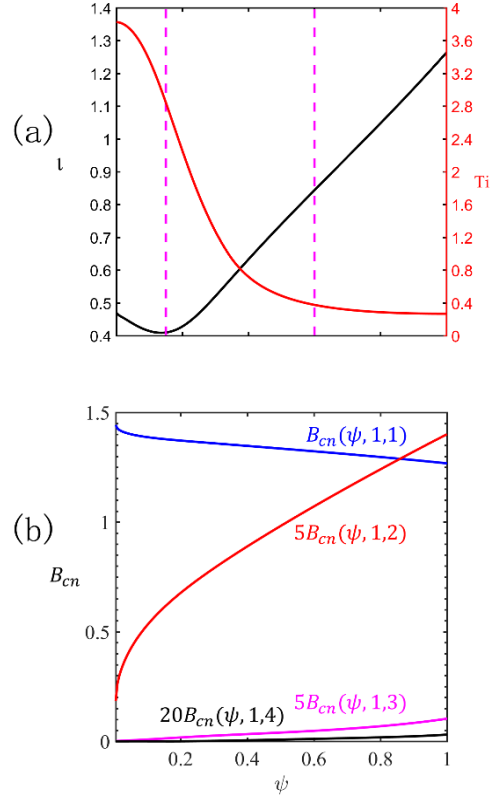

Figure 16. (a) Ion temperature (red) and rotational transform (blue) profiles. Simulation range is marked out by magenta lines: $\psi_{inner}$=0.15 and $\psi_{outer}$=0.6. (b) Fourier series coefficients to build magnetic field of LHD in GTC, $B_{cn}(\psi,1,1)$=0 (blue), $B_{cn}(\psi,1,2)$=5 (red), $B_{cn}(\psi,1,3)$=10 (magenta), $B_{cn}(\psi,1,4)$=15 (black). Values of $B_{cn}(\psi,1,2)$ and $B_{cn}(\psi,1,3)$ are multiplied by 5. Values of $B_{cn}(\psi,1,4)$ are multiplied by 20.

Given that the field period of LHD is 10, twice as large as the W7-X value, we expect that more parallel grids are needed. In our convergence study, the parallel grids are set to be 81, 121, 161. The growth rates and real frequencies calculated using these grids for mode `nmode=40, mmode=66` are given in Figure 17. For LHD, 121 grids in parallel direction are shown to be enough to get convergence in full torus simulations of LHD. According to the parallel grids convergence results for W7-X and LHD, as a rule of thumb, roughly 12 parallel grids are needed for every field period of stellarators in full torus GTC simulations.

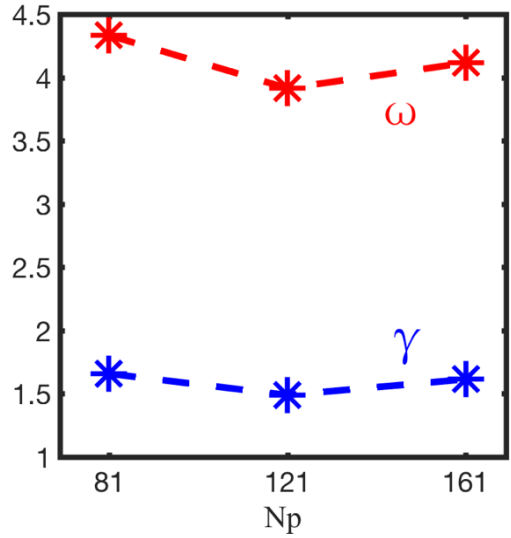

Figure 17. Growth rates (blue) and real frequencies (red) of parallel grids convergence.

The influences of modes coupling on growth rates and real frequencies in LHD are also studied. Applying the same analysis method used in W7-X, one nmode, 3 nmodes, 7 nmodes and all nmodes are kept in four cases by adding a filtering step. For `Nm=1` case, `nmode=40` is kept. For `Nm=3`, `nmode=30, 40, 50` are kept, and for `Nm=7`, `nmode=10, 20, 30, 40, 50, 60, 70` are kept. Growth rates and real frequencies of `nmode=40, mmode=66` are calculated and presented in Figure 17. Results show that growth rates and real frequencies of different nmodes coupling does not differ a lot in LHD, and the maximum value and the minimum value differs about 4%.

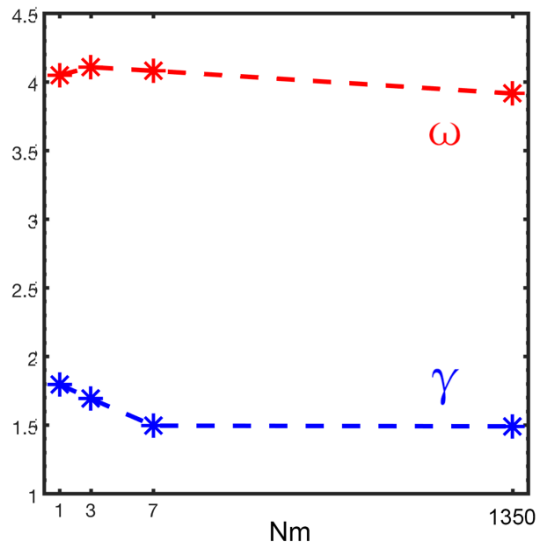

Figure 18. Growth rates (blue) and real frequencies (red) of Nmode convergence.

To study the most unstable modes in LHD, an average poloidal spectrum of diagnosis flux surface and the adjacent 10 flux surfaces is given in Figure 19. The most unstable mode is around `mmode=64` in LHD, which is lower compared to that in the W7-X case.

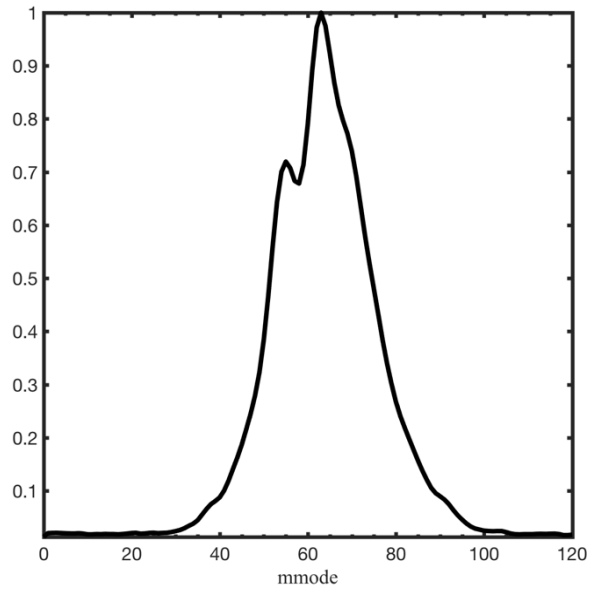

Figure 19. Averaging poloidal spectrum of different flux surfaces. X axis is the poloidal mode number, `mmode`.

The electrostatic potential on the diagnosis flux surface is shown in 2D in Figure 20, and in 3D in Figure 21, on top of the flux surface shape in real space. The potential intensity shows big difference between field periods, which is similar to W7-X, but is more obvious.

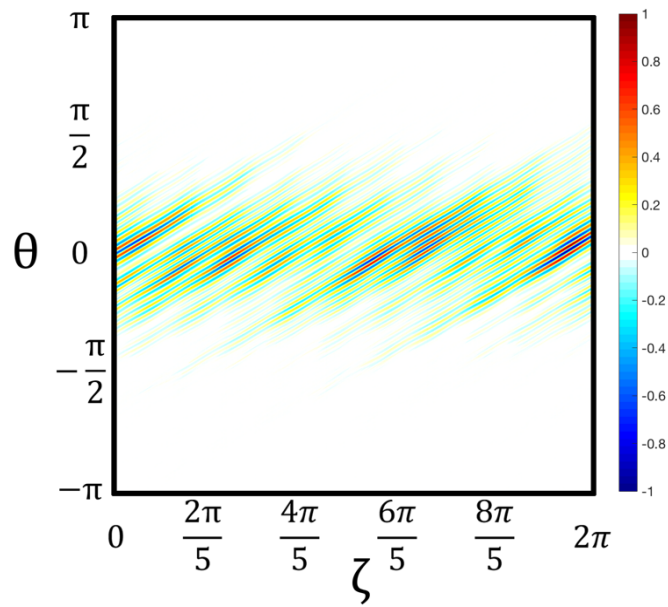

Figure 20. Phi on diagnosis flux surface. X axis is $\zeta$, from 0 to $2\pi$, and y axis is $\theta$, from $-\pi$ to $\pi$.

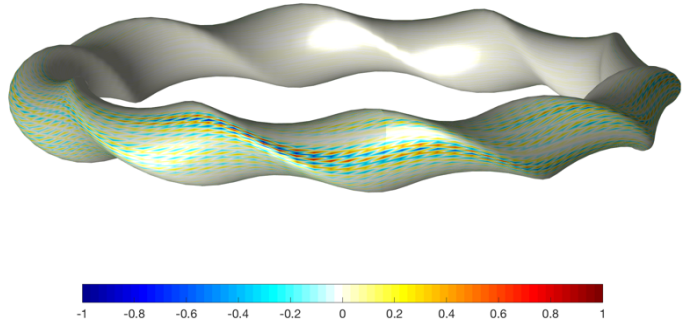

Figure 21. 3D electrostatic potential mode structure of LHD.

## V. CONCLUSION AND DISCUSSION

GTC is now able to simulate plasmas in non-axisymmetric equilibrium configuration. Simulation results of ITG driven instabilities in W7-X and LHD are given in Section Ⅲ and Section Ⅳ, respectively. The positive and linear correlation between the ion temperature and growth rate, as well as the real frequency, of the ITG mode in W7-X is observed and compared with the results of EUTERPE. The convergence in parallel grids is studied. A rule of thumb of the required parallel grid number has been developed, which helps better understand the feature of stellarators and provides a guidance for future works on stellarators using GTC. The effect of toroidal modes coupling in W7-X and LHD is shown to be important. According to the simulation results, the mode structure in W7-X is more localized, while LHD has a more global tokamak-like mode structure. The most unstable modes driven by ITG are different in W7-X and LHD. The toroidal mode number of the most unstable mode is significantly higher in W7-X than that in LHD. As a result, the mode structure is mode strongly affected by mode coupling in LHD than it is in W7-X.

We plan to use GTC to further study the electromagnetic instabilities, such as the Toroidal Alfven Eigenmodes, on stellarators.